# A cognitive approach to the topography of the 12<sup>th</sup> dynasty pyramids.

Giulio Magli
Faculty of Civil Architecture - Politecnico di Milano
Piazza Leonardo da Vinci 32, 20133 Milan, Italy

*The topography of the royal pyramids of the 12<sup>th</sup> dynasty is investigated in its historical and chronological context, in order to highlight connections between the architectural choices, the religious ideas and the traditions inspiring the reign of the corresponding pharaohs. A close, explicit connection with the "double" building projects carried out more than six centuries before by Snefru at Dahshur and at Meidum-Seila comes out. This connection involved geometry, perspective and astronomical alignments in the progressive planning of the pyramidal complexes. In such a context, the architectural choices of the last great king of the dynasty, Amenemhet III – who built two pyramids, one at Dahshur and the other at Hawara – appear to be due mostly to symbolic, as opposed to practical, reasons.*

## 1. Introduction

Scattered on the ridge of the desert between Dahshur and the northern rim of the Fayoum oasis, are the scant remains of the mudbrick pyramids of the Middle Kingdom Pharaohs. These pyramids were first studied at the end of the 19<sup>th</sup> century by De Morgan (1894) and Petrie (1902,1905). Later, they have been thoroughly re-excavated, so that today we have a very good knowledge of their architecture (Arnold 1987,1988, 1991, 2003; Stadelmann and Alexanian 1998).
Their ruined state is due to erosion and to the spoliation which occurred during the centuries. What we can see today does not, therefore, makes justice of the grandiosity of the original projects. Indeed, although they cannot be compared with the magnificent stone pyramids of the Old Kingdom, also the Middle Kingdom pyramids were conceived and built to be a visible symbol of power and to convey a series of messages related to the divine nature of the kings and their dynastic rights to kingship. It appears, however, that such messages have still, at least in part, to be understood. In particular, a key problem is the emerging role of Osiris in the funerary cult of the Middle Kingdom, and the contemporary decline of the "Heliopolitan" Sun God Ra in favor of Amun, the solar divinity coming from the heartland of the kings, Thebes. It is generally accepted that reflections of this evolving scenario can be seen in the evolving architectural choices of the pyramids' projects and, in particular, of their interior arrangements (Lehner 1995). However, several studies show that in the architecture of the pyramids of the *Old* Kingdom also the topographical choices had a fundamental importance to transmit ideas and relationships of religious and/or dynastic origin (Goedicke 2001, Jeffreys 1998, Lehner 1985, Magli 2009, 2010c, Verner 2002). This paper attempts to explore the same issue in the case of the Middle Kingdom pyramids within a "cognitive" approach (Flannery and Marcus 1996, Preucel 2010). The idea is thus to (try to) take into account all the aspects of human thought which can be seen reflected in the ancient topography. In particular, I will investigate here on the reflections of that peculiar mixture of "archaism and innovation" which pervades so many aspects of the Middle Kingdom (Silverman, Simpson and Wegner 2009).

## 2. The royal pyramids of the 12<sup>th</sup> dynasty: an overview

The 12th dynasty begins with the reign of Amenemhet I, who is also the builder of the first pyramid of the Middle Kingdom. We shall take into account here all the royal pyramids of this period (see map of sites in Fig. 1) and therefore I briefly report below, in chronological order, the kings, their accession dates (the delicate issue of co-regencies is not addressed) and a few basic information which will be needed in the sequel. As a reliable working framework, the chronology follows strictly that by Baines and Malek (1984) adopted by Lehner (1995), to whom the reader is referred to also for a much more complete survey.

1) Amenemhet I (1991 BC).
Amenemhet I was almost certainly born in Thebes, where one of his predecessor, Mentuhotep II, built the first great funerary complex of the Middle Kingdom located in the Deir el-Bahri bay (Arnold 1979). The reign of Amenemhet was certainly inspired by a close recall to ancient traditions and values, first of all the renaissance of the country as a unique kingship, as the Horus name shows (Wehem-mesut, "he who repeats births"). At the same time, we can see the rising importance of Amun (Amenemhet means "Amun is at the Head"). However, the throne name refers to Ra, as was the standard in the Old Kingdom and will remain in the future: Sehetep-ib-re, which means "Satisfied is the Heart of Re".
The king left Thebes (were an unfinished tomb was probably begun for him) and choose to found a new capital in the north. It was called Amenemhet-itj-tawy "Amenemhet the Seizer of the Two lands"; unfortunately, it has never been found, but it was probably located near the modern village of Lisht, were the king choose to construct his pyramid. The site is approximately mid-way between Meidum and Dahshur, the two sites were king Snefru constructed his magnificent pyramids some 600 years before. Although this specific fact is probably due to a chance, we shall see later on that the continuous reference to Snefru - thus to the epoch just before the time of the "solar" kings initiated by Khufu and kept to a maximum by the 5th dynasty kings – is a fundamental key to understand the Middle Kingdom pyramid's projects.
As in the Old Kingdom, also during the 12th dynasty each royal pyramid had its own name (referred only to the monument, no to the complex as a whole). For Amenemhet I it is "The Places of the Appearances of Amenemhet". The pyramid is located near the ridge of the desert, a rule always followed later on. The complex clearly resembles those of the Old Kingdom, with a valley temple, a causeway and a funerary temple located on the east side of the pyramid. Blocks coming from funerary complexes of the Old Kingdom, including Khufu's, were re-used in the pyramid's passages, but the bulk of the pyramid was built with mudbricks. The surface was then covered with slabs of white Tura limestone, the same used for Old Kingdom pyramids. Therefore, once finished, the aspect of the monument did not differ too much from that of these pyramids; the same holds for its height which probably was around 55 meters, comparable to Menkaure's at Giza. The entrance was in the north face at ground level, as was the standard in the 6th dynasty, with the chamber on the pyramid's vertical axis. However, as in Theban tombs, the chamber has a vertical shaft entrance; another similarity with tombs can perhaps be seen in the presence of two level terraces, one for the pyramid and one for the temple, and in the fact that the causeway was not roofed.

2) Senwosret I (1971 BC).
The funerary complex of this king is located at Lisht, about 1.5 Kms to the south of that of his father. The pyramid was called "Senwosret beholds the two lands" and included as much as 9 queens pyramids. Again, with an estimated height of 61 meters and a carefully laid casing of white stone, the pyramid was an imposing monument, constructed with the aid of a framework of radial walls. The causeway, originally un-roofed, was rebuilt with a roof and with the addition of a series of standing statues of the king in Osiris form, testifying the increasing importance of the Osiris cult. Indeed, it is in this period that this God, whose main center of cult was in Abydos, emerges as the main divinity connected with the underworld. This process is strictly related to the so-called "democratization" of the afterlife, which originally was prerogative of the living God, the king (Shaw 2000).

## 3) Amenemhet II (1929 BC)

With this king we enter in a new phase of pyramid construction (Arnold 1993).

In spite of the fact that the new pyramid's field established near the capital at Lisht was readily available for the construction of further monuments, Amenemhet II did not choose to build his pyramid (called "Amenemhet is Provided") near that of his father, but drastically changed the building site to Dahshur.

The Necropolis at Dahshur was already very old at that time. Since 600 years were standing there the two giant pyramids built by king Snefru (2575 BC). *No one knows* the reasons why Amenemhet II decided to leave Lisht and to choose Dahshur; for instance, in their authoritative surveys, Lehner (1995) does not discuss the issue at all, while Verner (2003) says that this choice happened "for some reason". Clearly, there *must* be reasons of symbolic, religious nature for such a choice.

In the Old Kingdom, the choice of the king's pyramid location reflected several criteria both of practical and symbolic origin (Barta 2005, Goedicke 2001). In several cases, the choice of the building site was indicative of the king's closeness to selected predecessors (think e.g. to Userkaf, who built his pyramid as close as possible to the Djoser complex in Saqqara, or to Niuserre, whose complex in Abu Sir is meticulously framed between those of his immediate antecedents). Further, explicit symbolism was embodied in the visual axes relating dynastically related king's monuments of Giza (Lehner 1985, Jeffreys 1998) and Abusir (Verner 2003) to Heliopolis, while in Saqqara Unas created one such axes to state his (ideal, besides actual) closeness to the Djoser complex, and in Saqqara south the $6^{th}$ dynasty pyramids were connected by meridian alignments to those of the Saqqara central field (see Magli 2010a, 2010b, 2010c for full details and a complete list of references). It is perhaps worth to put in evidence that nothing was "hidden" – or even worse "esoteric" - in this kind of topographical connections between monuments. On the contrary, the kings wanted to make explicitly visible their closeness to traditions, ancestors, or sacred sites trough the architectural features of their tombs.

In the case of Amenemhet II, a pretty similar mechanism must have occurred. To corroborate a symbolic, as opposite to functional, reason for the choice of Dahshur one can advocate the tale called *Prophecy of Neferti*. The *Prophecy of Neferti* is a text written during the early 12th dynasty (Simpson 1972). The story develops precisely at the court of King Snefru where a sage called Neferti is introduced to entertain the king. The pharaoh asks the sage to predict the future, and Neferti depicts a dark vision of the country in the hands of chaos. The depiction alludes to several misfortunes, including what could perhaps be interpreted as a total eclipse. Indeed, the text reads: "Re separates himself from men; he shines, that the hour may be told, but no one knows when noon occurs, for no one can discern his shadow...for he is like the moon in the sky". As is well known, the individuation of eclipses in ancient records is quite a difficult and delicate problem (see e.g. Baikouzis and Magnasco 2008); thus at a pure level of a curiosity I mention that a total eclipse actually occurred over Heliopolis during the first intermediate period, on 6/29/2159 BC (Julian). Chaos will, in any case, end with "a king from the south", called Ameny, who will restore order. The text has a propagandist aim, since "Ameny" is clearly Amenemhet I. Interestingly, the tale structure is close to that of the Papyrus Westcar, where a similar mechanism – a sage predicting a new generation of kings – is used to justify the rise of the $5^{th}$ dynasty. In the Westcar, the king "adopted" to officialize the new generation is Khufu. No doubt, the reign of Khufu represents an epochal breakthrough in the religion tradition and in the foundation of the divine power of the pharaoh. Khufu is indeed the first of the "solar" kings, and - at least according to authoritative scholars - he even depicted himself as Ra (Hawass 1993, Stadelmann 1991). In the Westcar, it is thus Khufu to absolve the duty of legitimating the new generation of kings, the "children of the Sun God" of the $5^{th}$ dynasty. In the case of the Neferti prophecy, the choice of the *last* great king before the solar tradition, Khufu's father Snefru, probably denotes the increasing distance between the pharaoh's "doctrine of power" and the solar, "Heliopolitan" ideas. Interestingly, the cult of Snefru in Dashour was extensively revived during the $12^{th}$ dynasty, while no such revival is documented at Giza (see Malek 2000 and references therein).

All in all, the choice of the building site of the pyramid of Amenemhet II at Dahshur has to be considered an explicit reference to Snefru. The monument lies to the east of Snefru's Red Pyramid and was surrounded by a huge rectangular enclosure oriented east-west. It is, unfortunately, in a completely ruined state due to the several despoliation occurred in the past, so that it is impossible to estimate its original height. Also the mortuary temple is almost completely destroyed, but its position can be individuated by the remains of two massive pylons on the east facade. The interior structure of the pyramid follows the standards which required the entrance to be in the middle of the north side. This will, however, be the last time for this kind of arrangement.

4) Senwosret II (1897 BC) ,
Senwosret II Khakeperre ("The Soul of Re comes into Being") decided to construct his pyramid at El-Lahun.
El-Lahun is located on the southern rim of the desert ridge just before the mouth of the Fayoum oasis channel. The choice of such place is usually explained with the "interest in the Fayoum oasis" by the king, who is credited to several drainage works in the area. Actually, it is certain that it was during the early Middle Kingdom that the Fayoum depression was reclaimed and transformed into a prosperous agricultural zone by diverting part of the Nile flow trough a channel leading to Lake Moeris [i]. It is however unclear if the end of these huge drainage works – perhaps started by Amenemhet I - has to be credited to Senwosret II; actually sometimes the presence of the king's pyramid at El-Lahun is advocated as the proof (see e.g. Shaw 2000). Perhaps the peculiar location of El-Lahun at the very "mouth" of the Fayoum channel and therefore in prominent position for all people entering the area, may be added to strengthen this connection.
The monument, originally around 50 meters high, is relatively well conserved, also due to the fact that it is constructed on an outcrop of yellow limestone (Fig. 2). The rest of the pyramid is of mudbricks with cross-walls inserted to help stability. The casing is lost today and was probably removed already at the time of Ramesses II.
The interior structure was investigated by Petrie (1891), who was initially unable to find the entrance. Indeed, the access is not located on the middle of the north side. It is instead in an impossible-to-guess position, namely in the pyramid courtyard near the east end of the south side. Clearly, concepts connected with the rebirth of the king in the circumpolar (northern) stars' region, which were mandatory in the Pyramid Texts of the Old Kingdom and consequently in the pyramid's architecture (Magli and Belmonte 2009), are lost here. Rather, it is the rising of the Osiris cult which influences the architectural choices. It has in fact being suggested that the "interest for the south" is due to the fact that Osiris' tomb was supposed to be located in Abydos, the main cult center of this God, and that the interior arrangement originated by analogy with the God's "apartment" in the underworld. Effectively, at the northwest corner of the burial chamber a passage loops around the room without any understandable function, and thus perhaps with a symbolic meaning; indeed the concept of "subterranean island" was associated with Osiris' death and resurrection. Another influence of the emerging cult can perhaps be perceived in the presence of a row of trees planted around the outer wall of the complex, possibly representing a grove supposed to be around Osiris' primeval tomb.
To sum up, the choices of this king are quite unusual and show what I believe to be a second, fundamental break in the traditions: first of all, the king changes again and abruptly from the pyramid field of his predecessor; second, its choice is a completely new site, third, the entrance passage is to the south, fourth, the apartments are "winding" towards the burial chamber.

4) Senwosret III (1878 BC).
The reign of Senwosret III can perhaps be considered the apex of the Middle Kingdom. This pharaoh is indeed credited of huge developments in the country and of consequent economic prosperity. His pyramid complex is again located at Dahshur, north-east of the Red Pyramid. However, as we shall see in a while, there is the strong possibility that the king was not buried there. In this case, the Dahshur pyramid would be a cenotaph, the word generically used to signify an empty tomb with a symbolic meaning. An example of a cenotaph dating to the Middle Kingdom is the so-called Bab el-Hosan, a tunnel – discovered by Howard Carter - located in the forecourt of the temple of Mentuhotep at Deir el-Bahri (Arnold 1979). It leads to a chamber which contained a seated statue of the king wrapped in line and an empty sarcophagus.

The pyramid of Senwosret III at Dahshur was a huge project, probably as much as 78 meters high (thus higher of all the pyramids from the 5$^{th}$ dynasty onward). Its entrance is located near the northwest corner. From there a vertical shaft drops to a descending corridor; the corridor then turns two times and arrives to a burial chamber which is quite out of the center of the pyramid. The chamber does contain a huge granite sarcophagus, decorated as a sort of miniature replica of the Djoser Step Pyramid enclosure wall at Saqqara. The room looks however to have never been disturbed or ravaged. No funerary equipment has been found, and no exploration has ever hinted at the existence of further royal chambers. Thus, the pyramid was perhaps not used for the king's burial; it was, however, used for the tomb of Queen Weret, as discovered by Arnold (2003). The funerary apartments of the Queen extend indeed under the satellite pyramid to the south and under the south-west quarter of the king's pyramid, with access from a shaft in the courtyard. Other queens and perhaps royal personages were buried to the north of the king's pyramid, in a underground cemetery which extends under the four satellite pyramids located to the north (Fig.6). Another relevant discovery made near the pyramid is that of six "funerary" wooden boats buried in the sand. Usually, this kind of finds is related to the solar cult tradition; however, it may be recalled that the early dynastic enclosures in Abydos were associated with burials of boats as well (O'Connor 2009). Actually, it is likely that Senwosret III was buried precisely in Abydos, were he built another funerary complex (Wegner 2007, 2009).

The complex is located some 2 Kms to the south of the Abydos center cult of Osiris. It is composed by a a funerary town, a mortuary temple near the cultivation and an extended underground tomb, whose entrance is in a T-shaped enclosure just below a conical hill, to the south of the temple. The axis of the complex is oriented towards this peak, an orientation of topographical origin. Indeed the peak is pyramidal in shape, so that in a sense we have here a first example of a tomb located under a natural pyramid, as of course will be the case in the New Kingdom with the Valley of the Kings located under the El Qurn peak of western Thebes. Joseph Wegner and his team found evidence that the mountain was devoted to Anubis and considered as sacred. Indeed, the seal of the Necropolis bears the text dw-Inpw *Mountain of Anubis.* Interestingly, the mountain-tomb is denoted with the symbolic hieroglyph of two paired peaks  , while in the New Kingdom (e.g. in Papyrus Abbott) the royal tomb will be denoted again with its "solar" ("Khufu") counterpart *Akhet*  .

The subterranean corridors of the Abydos complex are a masterpiece of Egyptian architecture. The tomb consists of two sectors, a first one employing limestone and a second one with chambers lined in red quartzite. After the main chamber, located on the axis, an arcing corridor contains two additional chambers. The curving corridor and chambers fall below the bulk of the pyramidal hill. The complex incorporates already many of the elements which will appear in the 18$^{th}$ dynasty burials of the Valley of the Kings. Basing on the analysis of the tomb's design, it appears that the tomb represents an archetypal expression of an "Amduat-tomb": a three-dimensional model of the 12-hour passage of the Sun God into the netherworld (Wegner 2009).

5) Amenemhet III (1844 BC)

Amenemhet III "Great of power" was the last great king of the Middle Kingdom. As that of his father, also the funerary project of this king comprises two monuments. This time, however, both monuments are pyramids, one in Dahshur and one in the Fayoum, and – contrary to what happens with Senwosret III - the reasons for this "doubling" of the project are usually considered to be purely functional. Following the tradition, I shall start here from the Dahshur pyramid.

The Pyramid of Amenemhet III at Dahshur is today called the Black Pyramid (Fig. 3). Although badly ruined, it is a quite imposing presence near the ridge of the desert to the east of the Bent Pyramid. The monument was probably 75 meters high, and its substructure is quite complex; it was first explored by De Morgan in a quite hastily way, and then thoroughly re-studied by Arnold

(1987). It comprises two apartments, internally connected by a corridor and usually denoted as king's and queen's sections respectively. The entrance to the king's section is located near the southeast corner on the east side of the pyramid, the entrance to the queen's section is located at the opposite end of a straight corridor, and therefore near the southwest corner on the west side. In the king's part, the burial chamber contains a pink granite sarcophagus, again with niches imitating the perimeter wall of Djoser's Step Pyramid. On the east side of the sarcophagus, symbolic eyes look in the direction of sunrise. The queen's section lies under the southern quadrant of the pyramid. There were two queens buried here, each one with her own burial chamber. The pyramid was violated in antiquity, but bones and a few items of funerary equipment were found in the queen's chambers.
To the south, a third, carefully designed system of chambers was built. It is accessed from both sections trough separate corridors and contains "ka" chapels which likely contained a statue of the deceased. The structure of this "south tomb" - which is unsuitable for burials – is very rigorous: to each burial chamber of the apartments corresponds a chapel, all the three being located quite precisely under the south side base of the pyramid. A transverse corridor runs further south, connecting these rooms.
According to Arnold, the pyramid was closed around year 20 of Amenemhet reign. It appears that most of the corridors were filled in with stones and mudbricks, as occurred in Senwosret III tomb in Abydos and will occur also in the other pyramid of Amenemhet III. Therefore, this filling was not due to the worries of the constructors about the possible collapse of the internal rooms. However, there is no doubt that the monument presented serious structural problems, as we shall see in more details in next section. This is usually taken as the reason for the construction of another pyramid by the same king.
The place chosen for this second pyramid is near the village of Hawara in the Fayoum. This place is *not* particularly favorable for a building site; to tell the truth it is a very bad position, being just a flat, relatively low land of desert. In any case, the pyramid – constructed, as usual, in mudbricks originally covered by fine limestone - is relatively well preserved (Fig. 4). The entrance is located on the south face near the southeast corner. Inside, the corridor leads north up to a death end. However, as in the Abydos tomb of Senwosret III, a corridor hidden within the ceiling leads, trough two other turns blocked by portcullises, first to an antechamber and then to the burial chamber. The chamber, located very near to the apex projection, is reported by Petrie to be a masterpiece of engineering (today the substructure is flooded by groundwater). It is composed by a "box" carved inside a huge quartzite monolith "sunk" into the chamber's pit. Inside lies the sarcophagus of the king, together with a smaller one. Huge quartzite slabs were lowered as ceiling of the room using a sand-lowering device. The whole chamber was protected by a saddle vault, located under a massive brick vault.

With the death of Amenemhet III the 12$^{th}$ dynasty approaches the end. There exists, however, two pyramids which might yet belong to the 12$^{th}$ dynasty, unfinished and in badly ruined state. They are located in Mazghuna, a few Kms south of Dahshur, and – basing especially on the similarity of Mazghuna south with the pyramid at Hawara - may belong to the successors of Amenemhet III, Queen Nefrusebek and Amenemhet IV. Their ownership is not certain however. Interestingly enough, two further unfinished mudbrick pyramids whose ownership is uncertain exist in Dahshur. One, located in Dahshur North, is probably a 13$^{th}$ dynasty pyramid, similar to that of king Khendier located slightly to the north. The other is reported to be located in Central Dahshur, south of Amenemhet II' pyramid; its scant remains were unfortunately much destroyed for the construction of a pipeline, but reliefs bringing the cartouche of an Amenemhet (number unknown) were found there.

## 3. The interpretation of the Amenemhet III project

Our analysis of the "cognitive" aspects of the topography of the 12$^{th}$ dynasty pyramids starts with the Amenemhet III project.
The accepted explanation for the construction of two pyramids into two completely different places is purely functional: it is thought that the Dahshur pyramid was considered unsafe due to its structural problems and it was decided to build a new pyramid at Hawara. However, there is practically no doubt that the construction of the Amenemhet III pyramid at Dahshur started during a co-regency of the two kings, when an enlargement of the Senwosret III complex was not yet realized (Arnold 2003). If we believe that the Senwosret III pyramid was designed only to be a cenotaph and a family burial, why the Amenemhet III pyramid, located less than 3 Kms far and probably designed by the same architects, was not?

Actually, as soon as one gets into the details of the "functional" explanation for the doubling of the Amenemhet III projects, problems arise. The main points are the following:

1) The Dahshur pyramid *was finished*. Why? Some scholars – such as Verner (2003) - adopt the view that the structural problems aroused "as soon as the monument was finished". This is clearly a *Petitio principii* : since the pyramid was finished, then the structural problems must have appeared after completion. Actually instead, from a technical point of view, it is impossible to assign a date to the crackings in relation to the progression of the building, because the *relative* progression between the construction of the substructure and that of the mudbrick mass is difficult to ascertain. What is clear is that, at a certain point, the weight of the building reached the limit of the subsoil resistance, which was perhaps affected also by the proximity to the ground water level (Arnold 1987). Once the limit of resistance per square meter was reached, the ground beneath the pyramid started to crack, and the substructure with it (it has to be excluded that it was the tunneling below the pyramid to be directly responsible for the cracking). As a consequence, extended fissures appeared and some of the corridors begun to collapse; in particular, crackings occurred in the ceilings of the rooms located near or directly under the south baseline of the pyramid. The architects had thus to face with such a dangerous situation, by reenforcing the corridors with wooden frames and mudbricks walls. Some of the emergency works were put in place in unfinished corridors, confirming that the dangerous events took place during construction. As a matter of fact, with such operations they interrupted the collapse of the interior apartments.
2) Some of the works carried out after the appearance of the crackings were clearly inspired by aesthetics rather than necessity. In particular, the basis of the king sarcophagus was carefully plastered with the aim of concealing the inhomogeneity of leveling due to the bulging of the floor.
3) The pyramid *was used as burial for the queens*. It makes little sense – at least to who writes – that the queens were to be buried in a structure considered unsure for the king.
4) The presence of unfinished or roughly finished corridors is a common feature of many of the Middle Kingdom pyramids, including Senwosret III'; therefore the collapsed status of some of them does not necessarily imply that the building was abandoned.
5) Crackings occurred also in the Senwosret III Queen's burial chamber, and were carefully repaired before burial. Therefore, the architects of Amenemhet III were well aware of the risk of constructing in proximity of the Nile floodplain.
6) It is difficult to establish when the works at Hawara started. Interestingly, carefully buried in the pavement of the annexes of the temple of the Dahshur pyramid, a limestone and wood model was found (Fig. 7). The model is quite accurate and certainly represents the interior arrangement of a pyramid, with a miniature wooden portcullis ready to be activated. It is pretty clear by a direct comparison, that this model represents the interior arrangement of the pyramid at Hawara. The unique difference is that only one sliding slab is present, instead of three. But taking into account the place were the model was found, it would be hard to sustain that it represents something else, like e.g. a yet undiscovered pyramid. Why was this model present in the center of the cult of the pharaoh in Dashour? The way it was buried resembles a deposit of foundation, and so the most likely answer is the existence of a symbolic, as opposed to functional, relationship between the two projects.
7) Finally, even admitting for a moment the "disaster" theory, why the building site for the new

pyramid was changed so drastically? Choosing a safer site nearby (e.g. the terrace about 1 Km to the south were the Pyramid of Ameny Qemau of the 13th dynasty will be later constructed) would have allowed the use of the already existing service structures.

To summarize, the Hawara pyramid was constructed for some reason, which does not seem to be due to the structural problems of the pyramid in Dashour. Why then? Actually, in the 60' of last century (before Arnold's new survey of the pyramid: De Morgan in his hasten exploration apparently did not realize the problems at all) Ahmed Fakry (1974) was writing about the Black Pyramid as follows: "There is no doubt that the king was buried in his pyramid at Hawara and that this pyramid, in the necropolis of the Old Kingdom Pharaohs, was a cenotaph". As a possible explanation for the choice of Dahshur for the cenotaph – instead of Abydos – Fakry mentions the fact that Snefru was deified and worshiped there during the 12th dynasty, and therefore the place was considered as sacred; an explanation worth considering for Senwosret III as well.
At the light of the above observations, Fakry's hypothesis seems to be the most viable. Furthermore, it looks very likely that the Dahshur pyramid was designed *from the very beginning* to be a cenotaph. It is indeed easier to think that the two projects, cenotaph in Dahshur and tomb in Hawara, were conceived together, because otherwise the complete difference in design between the internal apartments of the two pyramids would be inexplicable. The subterranean rooms in Dahshur are actually unique in design, while those in Hawara are very similar to those of Senwosret II in El-Lahun (Fig 5), and therefore cannot be considered as the result of a later change in funerary ideas occurred during the Amenemhet III reign.
Can "cognitive" aspects help in locating the 12th dynasty projects, including a supposed "global" Amenemhet III project, into a more coherent picture? In the next section I will re-run the "chronological topography" of the Middle Kingdom royal pyramids, with the aim of collecting their "cognitive" connections.

**4. The Snefru projects and the topography of the Middle Kingdom pyramids.**

As we have seen, Amenemhet II was the first king to return to Dahshur, and his choice was probably inspired by "archaism": the will of expressing closeness to the Snefru tradition.
A curious feature of the pyramid complex of Amenemhet II is that it is relatively small – the side base of the pyramid is estimated to be only 50 m. The smallness of the monument is enigmatic, also because that of this king was a prosperous reign. Perhaps, again, the explanation must be found elsewhere, in symbolic rather than functional reasons. Actually, the complex was located in a carefully chosen position with respect to the Snefru complex. If the line of the south base of the Red Pyramid is prolonged due east, it intersects a dense area of 4th dynasty tombs (Fig. 8). Immediately to the south of this area, we find the traces of the north side of the temenos wall of the Amenemhet II pyramid. It seems, therefore, that the latter complex was located in such a way to obtain a perspective effect with the much higher, but farthest in the desert, Red Pyramid of Snefru, thereby creating a visual – and symbolic – relationship.[ii]
The second pyramid constructed in Dahshur, that of Senwosret III, was planned to the north of that of Amenemhet II. The impressive artistic and symbolic value of the Senwosret III complex *per se* has been thoroughly discussed by Arnold (2003), and I will concentrate here on the topographical relationships with the other complexes.
A first topographical relationship holds with the Amenemhet II complex. This relationship, similar to those existing between the pyramids of the 6th dynasty in Saqqara south and the Saqqara central field, ideally connects the two complexes by means of a meridian (north-south) line which runs along the west side of the temenos wall of Senwosret III and along the front (east) side of the temenos wall of Amenemhet II.
A second relationship between Senwosret III and the already existing projects can be found using archaeo-astronomical analysis. At the time of the construction of the Bent Pyramid and of the Senwosret III pyramid the azimuths of the setting sun at the winter/summer solstice at Dahshur with a flat horizon were ~242°/298° respectively (sun azimuths do not depend on precession; they vary a bit due to the variation of the ecliptic's obliquity, so that today they are slightly displaced). The causeway of the Bent Pyramid is oriented (from the Valley Temple to the pyramid) at 240°. This means that, for an observer looking along the causeway, the Sun at the winter solstice was seen to disappear behind the huge mole of the pyramid and, as first noticed by Belmonte (2009), the Sun was seen to "lean" at the north-west corner of the pyramid by an observer positioned at the center of the Valley Temple, perhaps facilitating calendrical observations.

The architects who designed the causeway of the Senwosret III complex choose to create a configuration *symmetrical* to that designed more than 600 years before for Snefru. Indeed the causeway is oriented at 298°, and therefore points to the setting sun at the summer solstice. Since the pyramid is slightly to the north of the junction between the causeway and the temple complex, the mid-summer sun was seen to set framed between the south-west corner of the pyramid and the summit of the temenos wall. The name of the pyramid made reference to the *Ba* of the king as those of the middle 5[th] dynasty complexes did, perhaps emphasizing the solar connotations of the monument.

When the architects of Amenemhet III started the project of the king's pyramid, they apparently took into account the existing monuments in order to harmonize the new element in the human-made landscape and, one would be tempted to say, to keep *Maat*, the Cosmic Order, in the already old royal Necropolis;

1. First of all, the existing meridian was taken into account: it runs indeed along the west side base of the Black Pyramid.[iii] To fix the position of the pyramid *along* the meridian, the project took into account the position of the Bent Pyramid to the west, and again the new pyramid was planned in order to create a perspective effect between the new and the old one, as was the case for Amenemhet II and the Red Pyramid (Fig. 9).
2. It is difficult to ascertain the slope of the Black Pyramid, because its casing is almost completely lost. However, from remains found at the base it is estimated between 54° and 56°. If uncertainty about the slope cannot be removed at present, we are in any case certain of the slope of the pyramidion, which was recovered intact at the beginning of the last century and is now in the Cairo Museum, This slope is 54° 30'. Interestingly enough, this is *the same slope* of the lower courses of the giant "counterpart" of the Black Pyramid, that is, the Bent Pyramid.[iv]
3. A final clue to the "harmonization" of the king's project with the pre-existing ones is in the choice of the direction of the causeway, which runs due west. If we analyze the causeways at Dashour from south to north we see that their orientations obey the following "order": winter solstice sunset (Bent Pyramid), due west (Amenemhet III, Amenemhet I and – probably – the un-excavated Red Pyramid's causeway), summer solstice sunset (Senwosret III),

To sum up, then, we should try to imagine how the Necropolis of Dashour was looking like after the Middle Kingdom addictions. The landscape towards the Nile was quite different from now: lake Dahshur was extending to the south, in front of the desert ridge and of a (recently recovered) quay located in the wadi to the east of the Valley Temple of the Bent Pyramid (Alexanian et al. 2010). A visitor sailing the Dashour lake would have perceived a spectacular perspective effect, with the two imposing monuments of Snefru in the background and their respective companions, those of Amenemhet II and III, in the foreground. Aligned with these, further north, the imposing pyramid-cenotaph of Senwosret III was towering above the whole area.

The ideal resemblance to the old Snefru project in Dahshur can therefore be said to be complete with the completion of the Amenemhet III project there. If, as supposed by Stadelmann (1993) the Bent Pyramid was, if not conceived as part of a coherent project, at least finished as a cenotaph, then the resemblance would strengthen once again the interpretation of the Black Pyramid as a cenotaph as well.

We shall now investigate if cognitive aspects played a role also in the choice of the building site at Hawara. Indeed, Amenemhet III' choice of the Fayoum area is usually justified by the "interest" shown by the king for the oasis, where he built also a temple and, according to Petrie who discovered pieces of them, two huge quartzite colosses. However, as far as I know, the reasons for his option to the Hawara site are a sort of enigma which nobody ever dared to investigate. For instance, Verner (2003) seems to justify the choice because the place is "not far from Senwosret II pyramid in Lahun". Actually however, the two sites are 8.7 Kms far apart, and this is clearly *very* far from the point of view of a building site. At Lahun there was plenty of space to build a new pyramid, in an already existing - "sanctified" - necropolis of a revered predecessor. Further, there were the "infrastructures" needed for pyramid construction: accessibility of materials and a huge pyramid's workers town, Kahun, which remained active long after the reign of the king. So, once chosen the Fayoum area, why Amenemhet III moved from El-Lahun to the relatively remote and unsuitable site of Hawara?

First of all, it must be observed that the two pyramids of El-Lahun and Hawara are inter-visible. Today it is simpler to see the huge mass of El-Lahun from Hawara than viceversa, but in ancient times the two monuments clearly "spoke" with each other. They actually stand as "paired sentinels" at the two corner ends of the strip of desert which is the prolongation to the south of the pyramid's fields ridge. Observe now that, midway between Lahun and Lisht, stands the pyramid of Meidum, a project which was completed – if not initiated – by Snefru. Together with Meidum, the architects of this great king constructed yet another pyramid, that usually identified today as the "Seila pyramid" although the modern village of Seila lies quite far, in the Fayoum land plain. This small ("minor") step pyramid is located on a somewhat prominent desert hill to the west of Meidum, overlooking the Fayoum. The true function of such "minor step pyramids" - seven are scattered along the whole country, up to Elephantine – is unknown; they do not contain chambers, and perhaps it was calendrical (Belmonte, Shaltout and Fekri 2005, Belmonte and Shaltout 2006). In any case, the unique which is dated beyond doubts to the Snefru reign (due a stela found in the excavations, see Lesko 1988, Swelim 2008, 2010) is Seila, which is also the northernmost and the unique to be precisely oriented to true north. But why was it constructed? In the opinion of who writes, the main role of this small monument was of ideal companion of Meidum, which is much greater but located in the flat land, as "outpost cenotaphs" signaling the royal power in the approach to the capital, some 60 Kms further north. The two monuments indeed appear to be strictly related; today they are barely inter-visible with the naked eye, however in ancient times visibility was certainly better and the they are only 10 Kms apart (Fig. 10). Further, it seems that Seila was meant to be placed on the same parallel of Meidum. The azimuth of Seila from Meidum is indeed about 3° south of west. A deviation from the parallel of 3 degrees looks of course exaggerate for the Egyptian standards; it suffices to think to the orientation of the pyramids to the cardinal points, which was astonishingly precise in the Old Kingdom (Magli and Belmonte 2009). However, we should bear in mind that, while precise methods of meridian orientation using the motion of the stars around the celestial pole were available, no such methods existed for the alignment to a fixed point along the parallel (the parallel of the Meidum pyramid actually passes about 600 meters to the north of the Seila pyramid).

All in all, Meidum and Seila were *paired* monuments, mainly designed (or perhaps re-designed, as far as Meidum is concerned) to transmit a message of power. My proposal is that, in addition of ideally completing – together with the existing Amenemhet II complex - a ideal "replica" of the Snefru project in Dahshur with his first pyramid, the choice of Hawara allowed king Amenemhet III to ideally complete – together with the existing Senwosret II complex in El-Lahun – a second "replica": that of the Snefru project in Meidum. Indeed, if we look at the map (Fig. 1) we see that El-Lahun and Hawara play the role of "sentinels of power" in a pretty similar fashion to that played by Meidum and Seila. Seila is inter-visible with Hawara, and already Petrie (1891) noticed its prominent role and its connection with the Fayoum pyramids, a thing that actually led him to attribute Seila to the Middle Kingdom. He wrote indeed that Seila "is a landmark of all this part of the country; and can be seen from Hawara, as a white heap on the hill top...is probably a building of

the 12[th] dynasty"

## 5. Discussion

In a paper published in the seventies of the last century, Alice and Thom Kehoe wrote: "The archaeologist must approach his data with the expectation of describing concrete objects that in reality had their primary cultural existence as percepts in topological relation to one another within the cognitive schemata of human beings." (Kehoe and Kehoe 1973). Here, I have tried to follow such an approach treating as "objects" the 12[th] dynasty pyramids and analyzing their mutual topography. The latter can be better understood if symbolic, explicit connections of religious and dynastic nature between the various monuments are advocated. As already well known, the emerging role of the "southerly" God Osiris comes into play prominently; however, a key role turns out to be played also by "archaism" in the form of a close recall to the Snefru tradition and, specifically, to the topography of his own monuments.
In the opinion of the present author, a difficult topographical issue remains: to fully explain the ground-breaking choices of Senwosret II. He decided to change from the building site inaugurated by his dynasty, Lisht, and from the traditional site revived by his father, Dahshur. He broke the tradition of an orientation to the north of the entrance to the pyramid, and planned new, "winding" arrangement of the interior chambers. Both the successors of this king will have two tombs: Senwosret III had a pyramid tomb in the north at Dashour but was probably buried in the south one (at Abydos) and also Amenemhet III had a pyramid in the north at Dashour but was probably buried in the south, at the Hawara pyramid. Was also this tradition established by Senwosret II?
Basing indeed on the similarities between the internal arrangements of El-Lahun and Hawara, which lead to think that they are both "tombs to the south", one is also led to speculate that perhaps also Senwosret II might have had a cenotaph "to the north", where at least part of the *previous* traditions – building site and/or orientation - were still preserved.
This possibility has been suggested already by Arnold (2003) who noticed the peculiar structure of Petrie's Tomb 621, located to the north of the Lahun pyramid. This tomb – whose owner is unknown – has many peculiarities of a royal tomb; no burials remains have however been found in it and it may therefore be a cenotaph of the king; being "still" oriented to the north it might be the "missing ring" between the Amenemhet II and the Senwosret II pyramids.
To this possibility I would add, however, yet a different one.
The pyramid at Lahun was called, exactly as those of Snefru in Dahshur, "Senwosret shines". I would not be surprised if the unfinished Middle Kingdom pyramid at Central Dahshur, which has no owner, might one day be attributed to a cenotaph of Senwosret II, crediting him therefore as the first king of the 12[th] dynasty who had a "north" pyramid cenotaph in Dahshur and a "south" burial.

## Acknowledgments

The author gratefully acknowledges Juan Belmonte for several constructive comments to a first version of the present work.

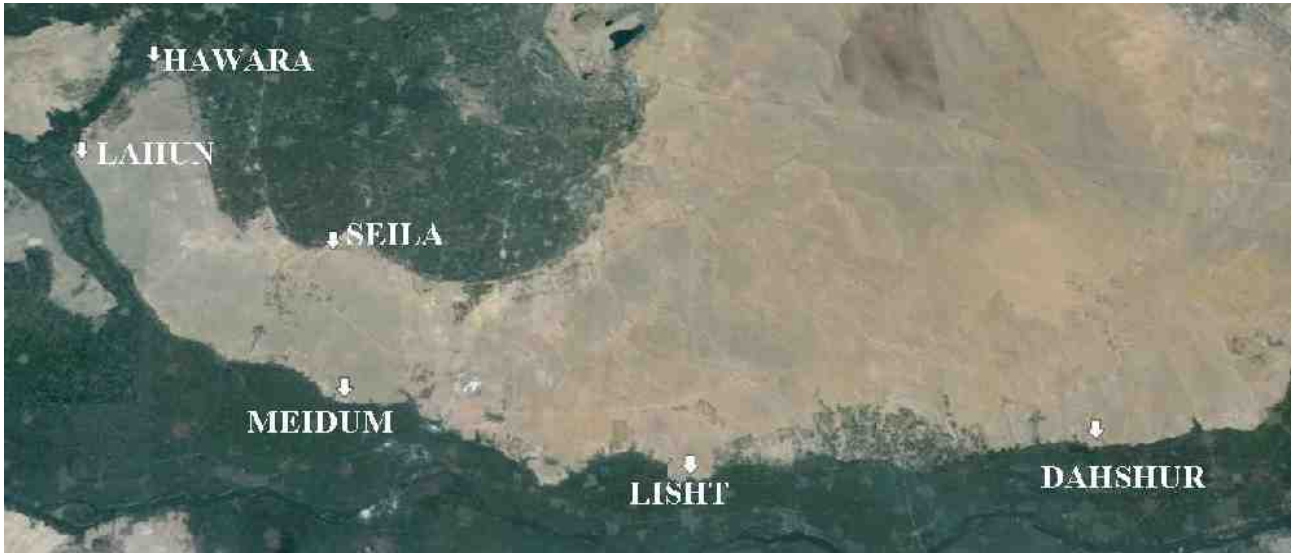

**Fig. 1. The archaeological sites cited in the text.**
**(North to the right; image courtesy of Google Earth, adapted by the author)**

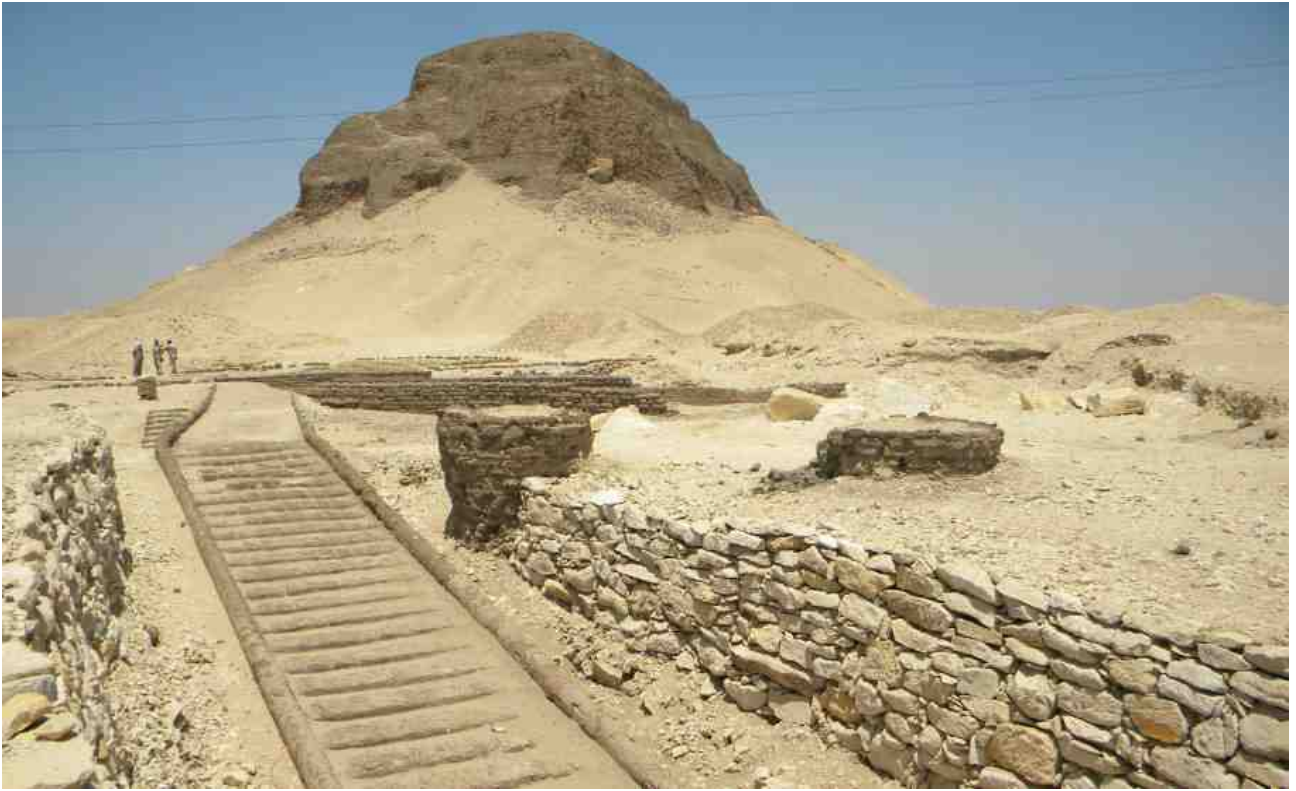

**Fig. 2. The pyramid at El-Lahun, view from the east (Photograph by the author).**

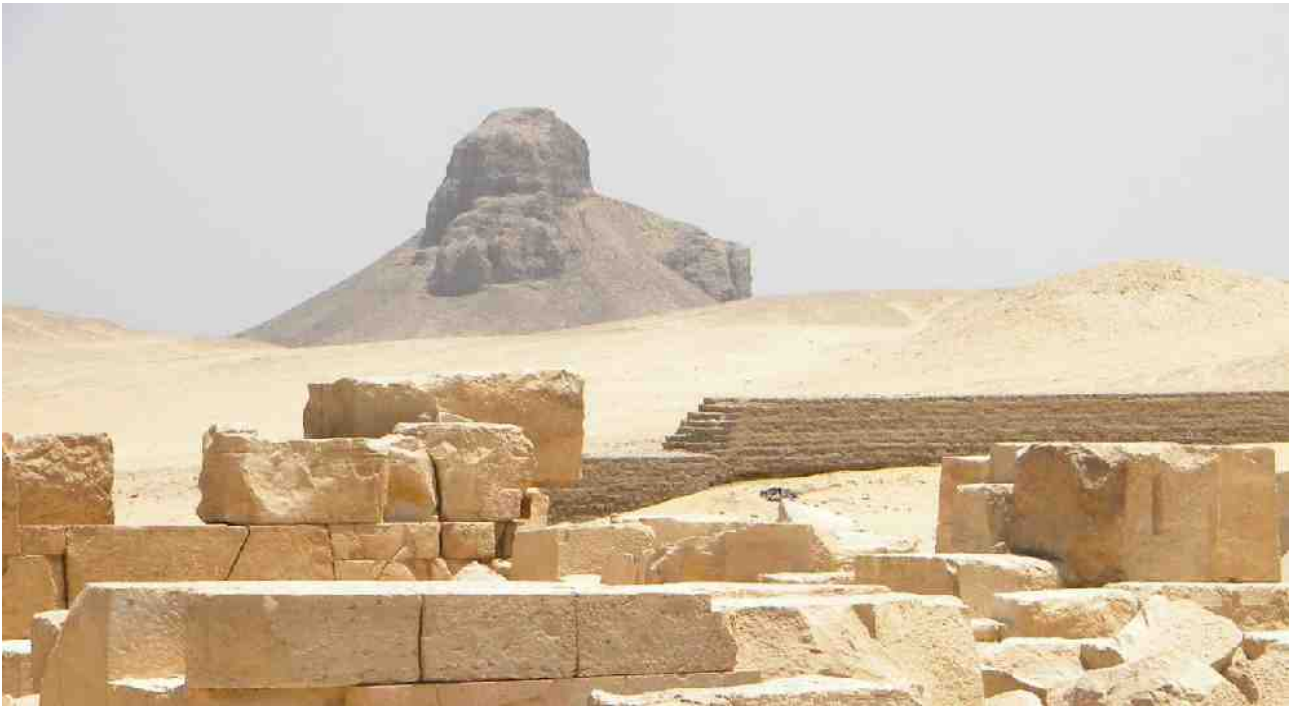

Fig. 3. The Black Pyramid at Dahshur seen from the Valley Temple of the Bent Pyramid (Photograph by the author).

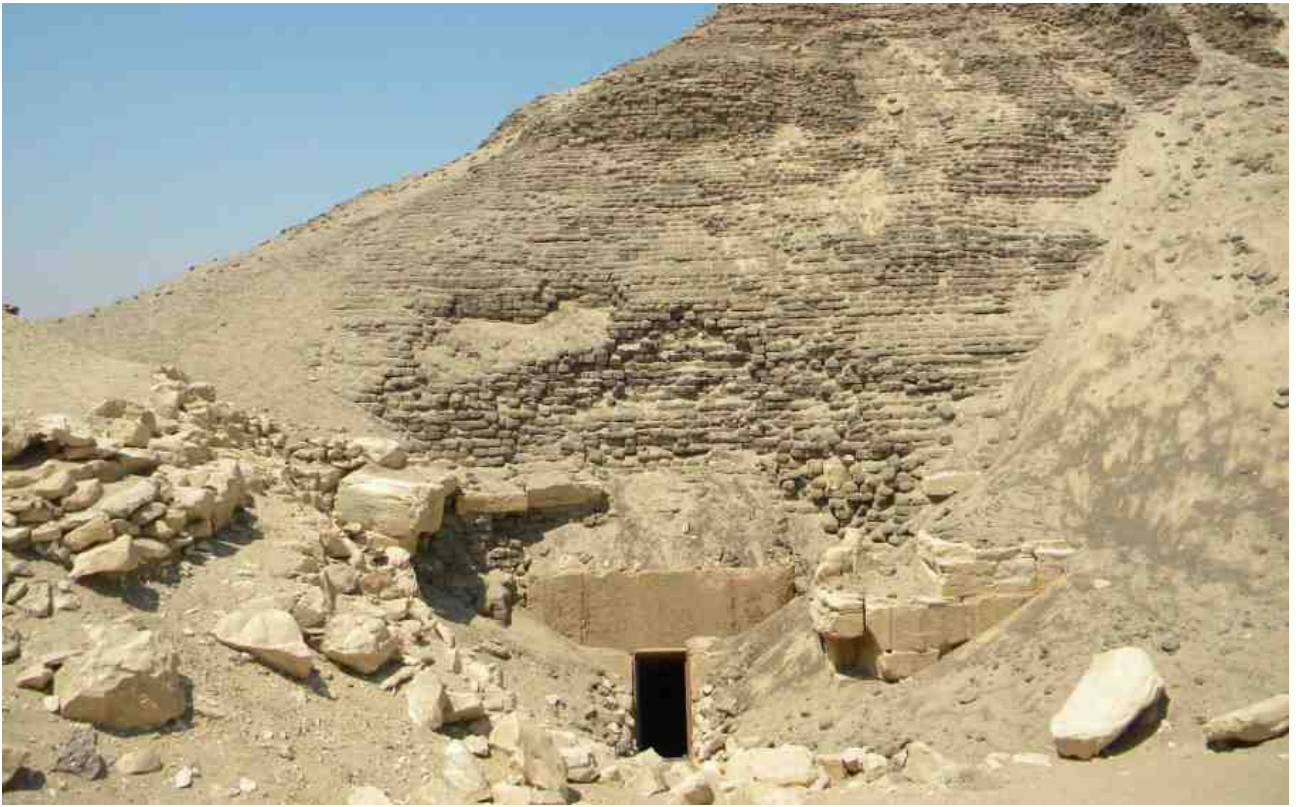

**Fig. 4. The pyramid at Hawara, the south side with the entrance (Photograph by the author).**

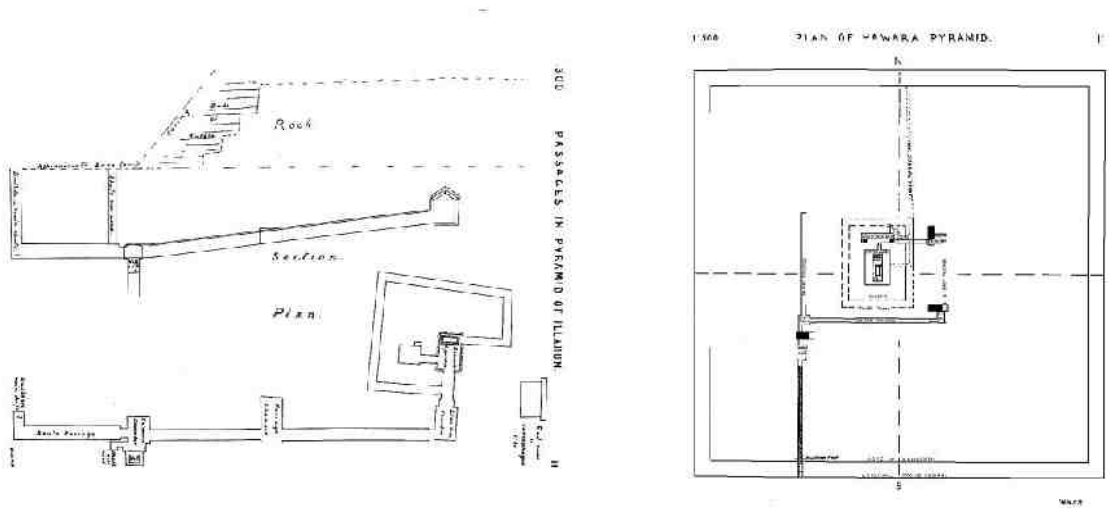

**Fig. 5. The substructures of the pyramids of Senwosret II at El-Lahun (left) and Amenemhet III at Hawara (right) (Adapted from Petrie; plans not at the same scale).**

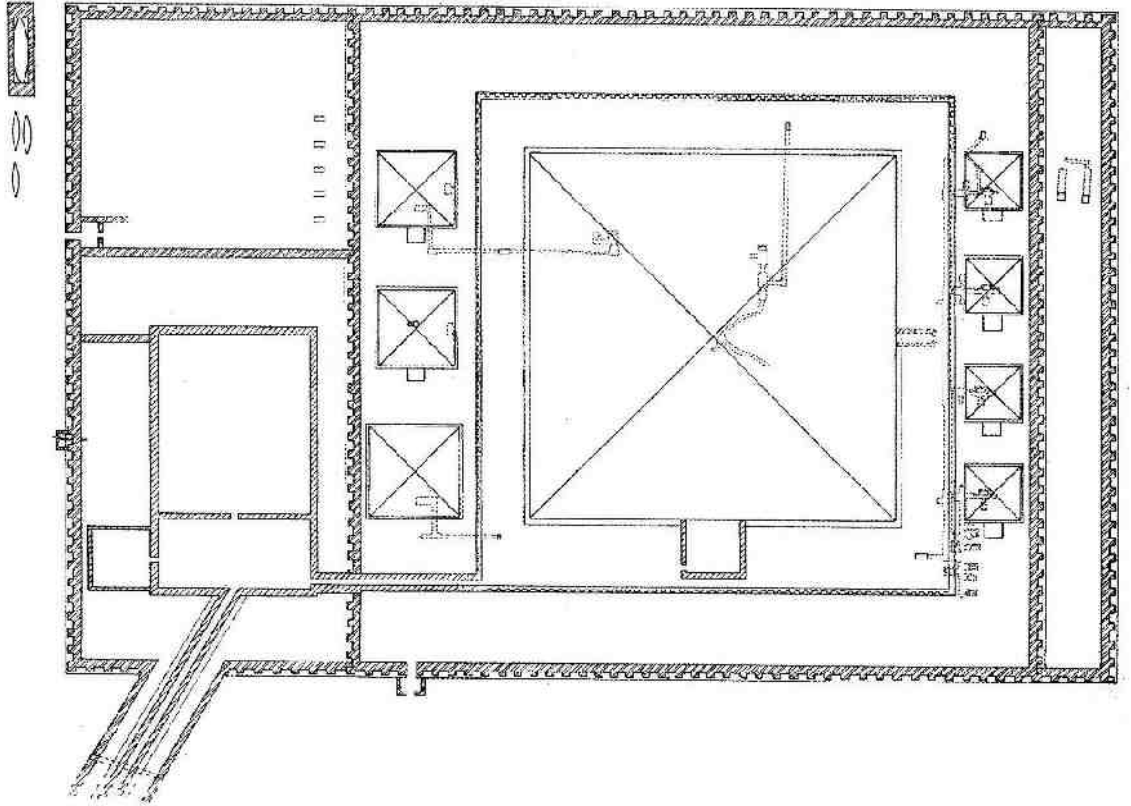

**Fig. 6. The complex of Senwosret III at Dahshur (Adapted from Arnold; north to the right).**

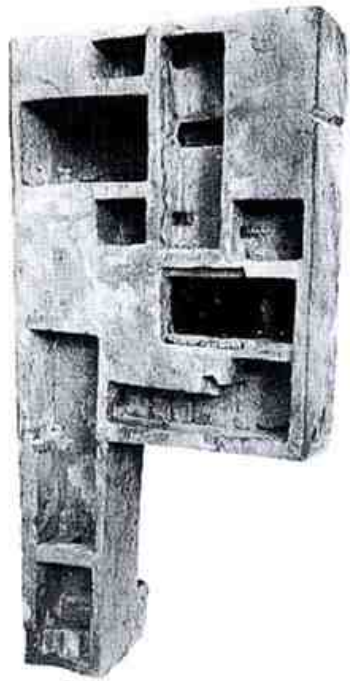

**Fig. 7 Model of the interior arrangement of a pyramid, probably that of Hawara, found in the Valley Temple of the Black Pyramid (compare with Fig. 5, right).**

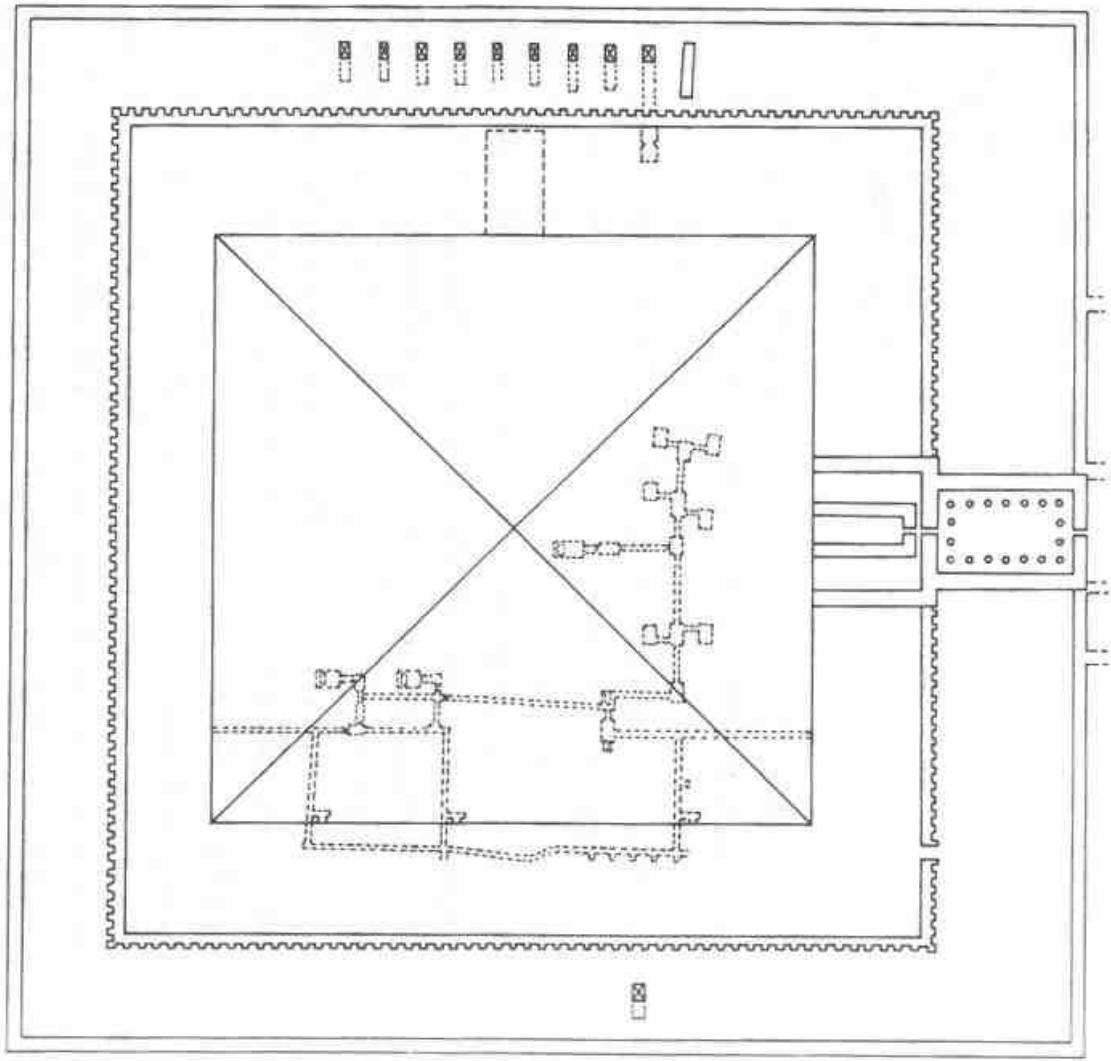

**Fig. 8. The substructure of the pyramid of Amenemhet III at Dahshur. (Adapted from Arnold; north on top).**

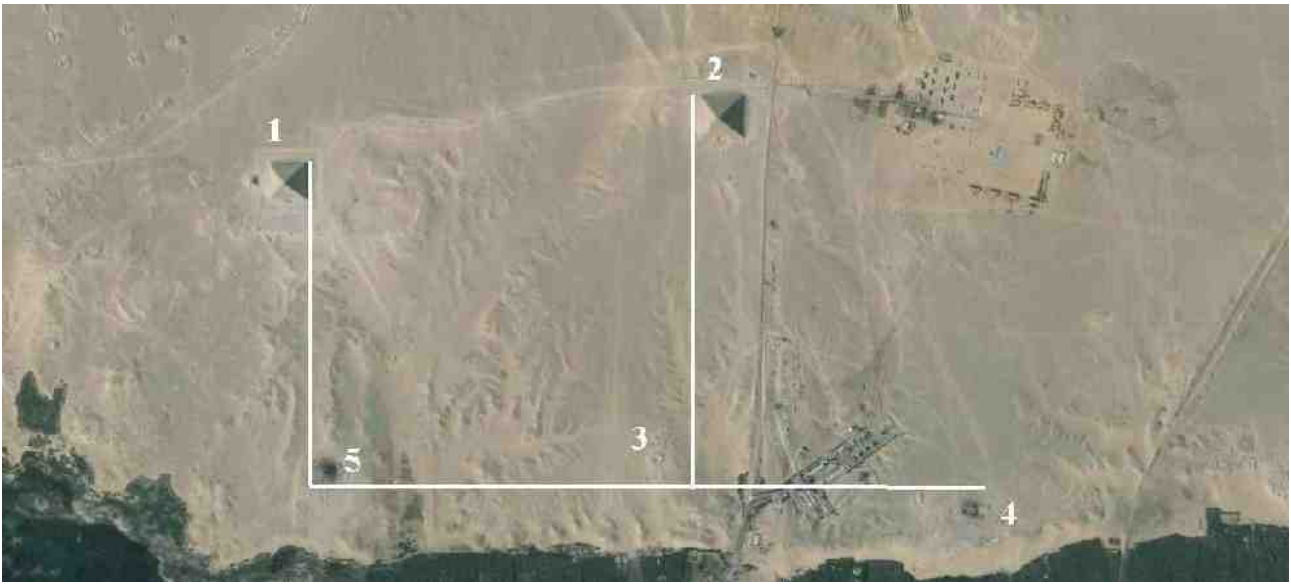

**Fig. 9. Air view of the Dahshur central field**
**(North on the right; numbering of the monuments in chronological order) 1) Bent Pyramid, 2) Red Pyramid, 3) Amenemhet II, 4) Senwosret III, 5) Amenemhet III. The reader should be warned that the white lines are drawn only as an help to the eye and might give a seeming impression of high precision.**
**(Image courtesy of Google Earth, drawings by the author).**

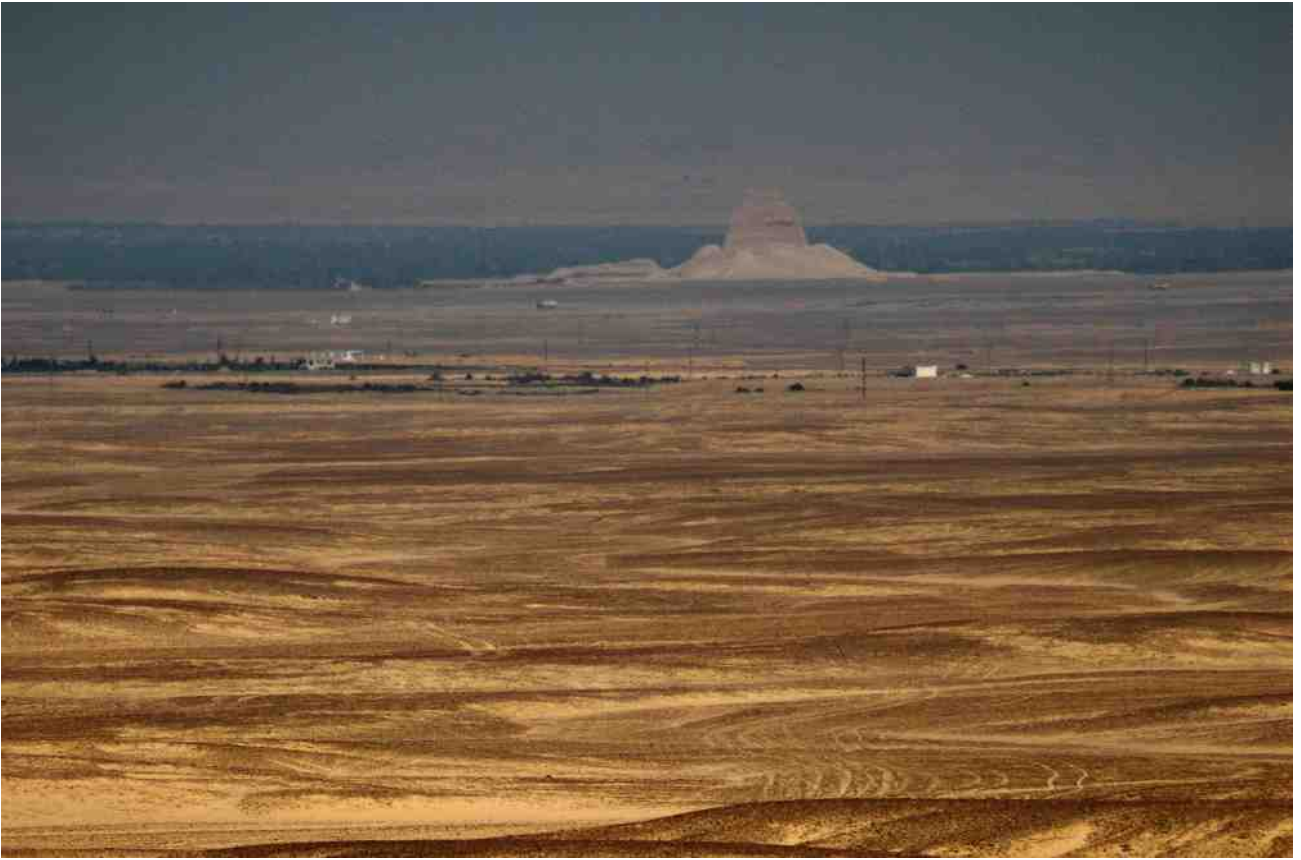

**Fig. 10. The Pyramid of Meidum viewed from the Pyramid at Seila
(Photograph by Juan Belmonte).**

<sup>i</sup> Interestingly, this was not the first great project of water regulation in Egypt; the first is probably the dam of Wadi el-Garawi near Helwan, about 32 kilometers south of Cairo, dated to the early 4th dynasty.

<sup>ii</sup> Besides the presence of the Old Kingdom necropolis, it would have been in any case not advisable to obstruct the view from the east constructing the king's pyramid exactly due east of the Red Pyramid - actually there is a unique example of such an intrusive design in all the history of the Egyptian pyramids, the Niuserre project in Abusir, which partially obstructed the view of the Neferirkare pyramid from the east.

<sup>iii</sup> If the same line is prolonged to the north, it runs also along the east wall of the 13th dynasty complex of Khendier in north Dahshur.

<sup>iv</sup> Also the pyramidion found in pieces at the base of the Red Pyramid (today reconstructed near the temple-chapel on its east side) has the same slope. It has been proposed that this pyramidion was prepared for the Bent pyramid before the decision of changing its slope (Rossi 1999); another possibility is, however, that the summit of the Red pyramid was capped with a pyramidion resembling in slope the lower courses of the Bent. Was it an attempt at "harmonization" already carried out at Snefru times? The *upper* courses of the Bent pyramid indeed actually have the same slope of the Red pyramid's courses, 43° 22' (Lehner 1995). To gain a complete picture it would be desirable to recover at least one casing block of the Amenemhet I pyramid, in order to ascertain its slope.